\documentclass[preprint,aps]{revtex4-1}
\usepackage{graphicx,color}

\begin{document}
\title{Flow and air-entrainment around partially submerged vertical cylinders}
\author{Valentin Ageorges}
\author{Jorge Peixinho}
\altaffiliation[Also at ]{Laboratoire PIMM, CNRS, Arts et M\'{e}tiers, CNAM, H\'{E}SAM Universit\'{e}, Paris, France} 
\author{Ga\"{e}le Perret}
\affiliation{Laboratoire Ondes et Milieux Complexes, CNRS et Universit\'{e} Le Havre Normandie, 76600 Le Havre, France}
\date{\today}
\begin{abstract}
In this study, a partially submerged vertical cylinder is moved at constant velocity through water, which is initially at rest. During the motion, the wake behind the cylinder induces free-surface deformation. Eleven cylinders, with diameters from $D=1.4$ to 16 cm, were tested at two different conditions: (i) constant immersed height $h$ and (ii) constant $h/D$. The range of translation velocities and diameters are in the regime of turbulent wake with experiments carried out for $4500<Re<240 \,000$ and  $0.2<Fr<2.4$, where $Re$ and $Fr$ are the Reynolds and Froude numbers based on $D$. The focus here is on drag force measurements and relatively strong free-surface deformation up to air-entrainment. Specifically, two modes of air-entraiment have been uncovered: (i) in the cavity along the cylinder wall and (ii) in the wake of the cylinder. A scaling for the critical velocity for air-entrainment in the cavity has been observed in agreement with a simple model. Furthermore, for $Fr>1.2$, the drag force varies linearly with $Fr$.
\end{abstract}
\maketitle
\section{Introduction}

The flow past ships \cite{rabaud2013,moisy2014}, or an emerged body such as bridge pillar, is a fundamental, familiar and fascinating sight. Measurements and modeling of this simple flow can have relevance for the floating offshore structures and renewable energy systems \cite{drew2009,lopez2013}. The disordered motion of the free-surface in the wake of the cylinder is related to the drag and lift effect experienced by the cylinder. The wake pattern is responsible for the surface waves and probably the air-entrainment. Many experiments and numerical works have considered a single vertical cylinder geometry in fully submerged situation \cite{schewe1983,braza2006,achenbach1981,breuer2000,tritton1959}. These works reported the drag force measurements on the circular cylinders as a function of the Reynolds number $Re$ defined as $Re=UD/\nu$ with $U$ the velocity, $D$ the cylinder diameter and $\nu$ the kinematic viscosity of the fluid. In addition, Schewe \cite{schewe1983} showed the occurrence of a drag crisis at $Re \approx 350~000$, which separates two distinct regions of nearly constant dimensionless drag coefficient, $C_D$, defined later. In the subcritical regime, for $ Re < 350~000$, the drag coefficient and the vortex-shedding frequency $f$ characterized by Strouhal number $St=fD/U$, are almost constant at $C_D\approx 1.2$ and $St\approx 0.2$. In this regime, the flow is characterized by a turbulent wake with a laminar boundary layer separation \cite{sumer2006}. 

In contrast to fully submerged situations, few studies have been done taking into account the effect of a free-surface, which modifies the pressure distribution around the cylinder. From a numerical point of view, Refs \cite{kawamura2002,yu2008,koo2014} have computed the flow around a wall mounted cylinder, the evolution of $C_D$ and the free-surface elevation for $Re$ varying from 27 000 up to 458 000. They showed that the vorticity and $C_D$ vary with depth along the cylinder and diminish close to the free-surface. Additionally, $C_D$ is strongly influenced by free-end effects for finite height cylinders \cite{sumner2004}. Specifically,  Gon\c{c}alves  {\it et al.} \cite{goncalves2015} reported experimental results in a recirculating water flume for cylinders piercing the free-surface with aspect ratio  $0.1<h/D<2$, where $h$ is the immersed cylinder height. They focused on forces and Particle Image Velocimetry (PIV) measurements for $10\,000<Re<50\,000$ and $0.07<Fr<0.36$; the Froude number $Fr$ is defined as $Fr=U/\sqrt{gD}$ with $g=9.81$ m/s$^{2}$ the gravitational acceleration. For constant $Re$, $C_D$ values increase with $h/D$. Moreover, the velocity fields at $Re=43\, 000$ revealed two  recirculation regions: (i) below the cylinder and (ii) behind the cylinder. The measurements in Chaplin and Teigen \cite{chaplin2003} also showed that $C_D$ values reach a maximum with respect to $Fr$. The evolution of $C_D$ values is in agreement with the experiments of Ducrocq {\it et al.} \cite{ducrocq2017}, who investigated the behavior of subcritical and supercritical flows for different slopes of a flume with $0.5<Fr<2.3$ and $Re$ = 50 000.

In the recent years, numerical studies have reproduced the flow and drag around free-end cylinders for different $h/D$ \cite{yu2008,koo2014,benitz2016,ducrocq2017}, but there is a lack of data for $Re$ around the drag crisis ($Re \approx 350~000$) and almost no data for strong free-surface deformations and rupture. An important numerical output is the mean interface elevation \cite{yu2008,koo2014}. Specifically, Keough {\it et al.} \cite{keough2016} analyzed the evolution of the fountain height ahead of the cylinder for $Re$ varying between 45 000 and 380 000. They concluded a dependency of the fountain height with $Fr^{2}$, and revealed that this evolution follows Bernoulli equation for small velocity. In addition, at the downstream side of the cylinder, the maximum cavity depth $\mathcal{L}$ is also believed to scale with $Fr^2$ \cite{chaplin2003}. The cavity depth $\mathcal{L}$, is schematized in Fig. \ref{fig::fig1}(c). Depending on cylinder diameter, the free-surface downstream is locally subjected to strong deformations and air-entrainment in the liquid phase. Benusiglio \cite{benusiglio2013} recently demonstrated a velocity threshold for air-entrainment in the cavity of a translating vertical cylinder. This phenomenon can induce drag reduction effects and it is independent of the drag crisis introduced before. At larger $Fr$, Hay \cite{hay1947} reported an extensive study of the ventilated regime, where air reaches the bottom of the cylinder and creates a ventilated cavity with large waves and sprays above the nominal free-surface. Modern studies of the artificially or atmospherically ventilated structures (plates or hydrofoils) close to a free-surface are concerned with the stability of ventilated cavities (e.g. Ref. \cite{harwood2016}) or properties such as the void fraction measurements \cite{makiharju2013}.\\

The goal of the present investigation is to quantify the onset of air-entrainment effects. In addition, the present study adds new experimental results and discusses new drag measurements carried out for $4\, 500<Re<240\, 000$ and $0.2<Fr<2.4$ with specific focus on the onset of air-entrainment \cite{kiger2012,chirichella2002,peixinho2012,harwood2016}. This paper is composed of two parts. In the first part, the experimental setup is described together with an optical method for free-surface height reconstruction. In the second part, results are presented with the drag force measurements and the determination of the velocity threshold for air-entrainment in the cavity. Finally, the air-entrainment results are compared with a simple model based on the balance between acceleration and gravity.

\section{Experimental setup}

\begin{figure}
\center
\includegraphics[width=1.0\columnwidth]{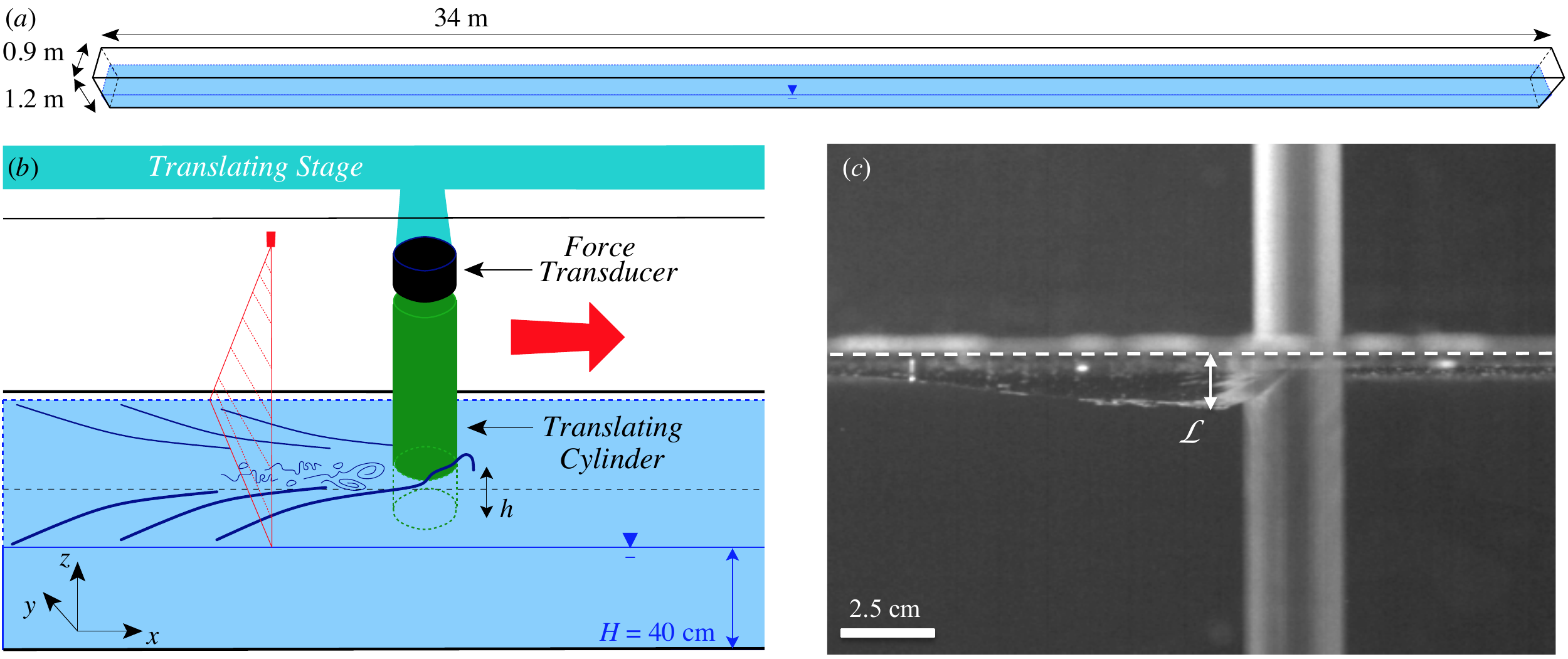}
\caption{Experimental setup. (a) Schematic tilted view of the flume (drawn up to scale) with the dimensions: $34\times1.2\times0.9$ m. (b) Schematic of the arrangement around the cylinder, view from the side above-surface. The shaded red area represents the laser sheet used for the optical method. (c) Typical picture side view. The cylinder is 2.5 cm in diameter and has been translated from left to right with a velocity of 45.8 cm/s, $Re=11\,500$ and $Fr=0.9$. The depth of the cavity $\mathcal{L}$ is represented in the near cylinder wake.}
\label{fig::fig1}
\end{figure}

The cylinders are hollow and made of PMMA with external diameters of 30, 40, 50, 60, 70, 80, 90, 110 and 160 mm. Their total height is 65 cm and their wall thickness is 5 mm. The Young modulus of PMMA is 3.1 GPa. There are additional hollow cylinders of 14 and 25 mm made respectively of copper and aluminum. All the cylinders are clogged at their ends. These cylinders are attached vertically, perpendicular to the water free-surface, and translated on a carriage riding horizontally along the flume of 34 m length, 90 cm width and 120 cm height. The flume is shown schematically in Fig. \ref{fig::fig1}(a). The carriage (162 kg) is driven by a 600 V AC motor (34 kg), producing a maximum torque of 102 N.m. A  protocol is applied where first the velocity increases linearly along a distance of 3.5 meters. Then, the speed is maintained constant over a distance of 9.5 meters, where force measurements are acquired in steady state. Finally, the speed decreases linearly over a distance of 2 meters.  

Two windows are located on the sidewalls of the flume, allowing local flow visualization around the cylinder. The determination of the critical velocity for  air-entrainment behind the cylinder is based on images from a 2000 $\times$ 2000 pixels CCD camera (Basler aca2040) with focal length of 50 mm placed at 6 m from the start of the cylinder run. At this position along the flume, the force measurement (described later) has reached a steady state value. The spatial resolution of the CCD camera is 300 $\mu$m per pixel and its acquisition frequency is 100 frames per second. Hence, bubbles smaller than 300 $\mu$m can not be detected. A typical image of the cylinder passing through the field of view of the camera is shown in Fig. \ref{fig::fig1}(c).

The working fluid is tap water and its height in the flume is constant: $H = 40\pm 0.1$ cm. The temperature is monitored hence the viscosity, the density, $\rho$, and the surface tension, $\sigma$, of water are corrected to take into account these variations. Indeed between all experimentations, the temperature of water changes from 16.1 to 21.3$^{\circ}$C. In this range of temperature, density variation represents 0.1\% \cite{tanaka2001}, while viscosity variations represent 13.3\% \cite{huber2009}. The correction allows us to keep the error for $Re$ between 0.8 to 9.8\%. The Bond number \cite{aristoff2009}, $Bo= \rho g D^2/\sigma$, varies between 26 and 3450, which implies that the capillary effects are weak.

Two series of experiments were performed: (i) at constant immersion depth $h=23 \pm0.1$  cm leading to $h/D$ ratio varying between 1.44 to 16.5 and (ii) at constant ratio $h/D= 2.55 \pm 0.05$. In total, 372 runs were performed, as summarized in Table \ref{tab:tab1}, covering a wide range of $Re$ and $Fr$. The maximum blockage width of the experiment, $D/W$, where $W$ the flume width, is 0.177, a value below the critical blockage width of 0.2 for confinement effect on three dimensional dynamics proposed by Griffith {\it et al.} \cite{griffith2011}. In addition, the ITTC 2002 \cite{muller-graf1987} requires a vertical confinement, $(H-h)/h>0.8$. For the current experiments, at constant $h$, the vertical confinement is 0.74 and for constant $h/D$, this criteria is satisfied for $D<9$ cm.

\begin{table*}
\caption{\label{tab:tab1} Range of $Re$ and $Fr$ numbers for constant $h$ and constant $h/D$ runs}
\begin{ruledtabular}
\begin{tabular}{cccccccccccc} 
~ & \multicolumn{5}{c}{$h=23\pm0.1$ cm} & \multicolumn{5}{c}{$h/D=2.55\pm0.05$} \\ 
\hline
$D$ (cm) &$h/D$ & $Re_{min}$ & $Re_{max}$ & $Fr_{min}$ & $Fr_{max}$ & & $h$ (cm) & $Re_{min}$ & $Re_{max}$ & $Fr_{min}$ & $Fr_{max}$\\
 1.4 & 16.50 & 7\, 470 & 12\, 460 & 1.45 & 2.41 & & 3.6 & 4\, 340 & 12\, 160 & 0.87 & 2.41\\
 2.5 & 9.20 & 12\, 460 & 24\, 910 & 1.00 & 2.10 & & 6.4 & 9\, 730 & 21\, 880 & 0.81 & 1.82\\
 3  & 7.77 & 17\, 440 & 29\, 900 & 1.08 & 1.84 & & 7.6 & 15\, 950 & 27\, 900 & 0.98 & 1.72 \\
 4  & 5.75 & 24\, 900 & 39\, 860 & 1.00 & 1.60 & & 10.2 & 19\, 930 & 39\, 860 & 0.80 & 1.60 \\
 5  & 4.60 & 30\, 630 & 59\, 800 & 0.86 & 1.71 & & 12.8 & 28\, 460 &  54\, 550 & 0.86 & 1.64\\
 6  & 3.83 & 40\, 840 & 71\, 460 & 0.87 & 1.52 & & 15.3 & 29\, 180 & 68\, 080 & 0.65 & 1.52\\
 7  & 3.28 & 29\, 900 & 97\, 260 & 0.52 & 1.74 & & 17.9 & 32\, 820 & 97\, 260 & 0.58  & 1.72\\
 8  & 2.87 & 34\, 040 & 122\, 510 & 0.50 & 1.70 & & 20.4 & 34\, 040 & 116\, 710 & 0.49 & 1.69\\
 9 & 2.55 & 34\, 880 & 119\, 600 & 0.41 & 1.42 & & 23.0 & 34\, 880 & 119\, 600 & 0.41 & 1.42\\
 11 & 2.09 & 39\, 860 & 188\, 870 & 0.35 & 1.62 & & 28.1 & 39\, 860 & 184\, 370 & 0.35 & 1.62\\
 16 & 1.44 & 40\, 860 & 239\, 190 & 0.20 & 1.20 & & 39.9 & 39\, 860 & 239\, 190 & 0.20 & 1.20\\
\end{tabular}
\end{ruledtabular}
\end{table*}

\subsection{Force measurements}

A piezoelectric sensor (Kistler 9327C), placed above the cylinder (see drawing  in Fig. \ref{fig::fig1}(b)) measures the axial force $F_x$ acting on the whole cylinder with a sensitivity of -7.8 pC/N. The sensor is connected to a charge amplifier equipped with a low pass filter of 30 Hz. The contribution of air drag is neglected. Indeed, when $h$ is minimum, the maximum error on $C_D$ is about 0.8\%. Force signals are filtered using a moving average on 20 points. In order to measure the effect of air-entrainment on drag forces, a time average is performed on $F_x$ during 5 s at constant translating velocity. It is denoted $\bar{F_x}$. The drag coefficient $C_D$ can thus be defined as: 
\begin{equation}
C_D(t) = \frac{2\bar{F_x}}{{\rho}{hDU^2}}. 
\label{eq1}
\end{equation}

The Strouhal number, defined earlier, is found to be close to $0.2\pm 0.025$ for $D<6$ cm, using the peak of $f$ from the spectrum of lift force signal. However, for larger diameters, the signal of lift force is noisy, presumably due to the turbulent flow or vibrations of the carriage. Moreover, the range of $Re$ studied here and the materials used here suggest natural frequencies which are well separated from those associated with vortex-induced vibration, which is the motion induced on the elastic cylinder interacting with the turbulent fluid flow. Such vortex-induced vibration phenomena of the elastic cylinder deformation would act like an oscillator inducing several modes corresponding to different vortex sheddings, observed by several authors \cite{fujarra2001,sarpkaya2004,williamson2004,franzini2012}.

\subsection{Optical method}

The parameter that needs to be measured is the elevation of the free-surface. The height at rest $H$ is well controlled. However, when the cylinder passes through the observation area, the light from a vertical laser sheet located above the flume and pointing to the free-surface, represented in red in Fig. \ref{fig::fig1}(b), is shifted vertically from its initial position. The shift is the difference between the interface at rest and the deformed free-surface. It is proportional to the elevation of the free-surface. The images are captured by a camera placed on the top  of the flume with a low incident angle to the free-surface. The proportionality is calibrated by recording images of the laser sheet at five known positions over 25 mm. This technique is able to detect free-surface heights with a precision estimated to $\pm0.1$ mm.

\begin{figure}
\center
\includegraphics[width=1.0\columnwidth]{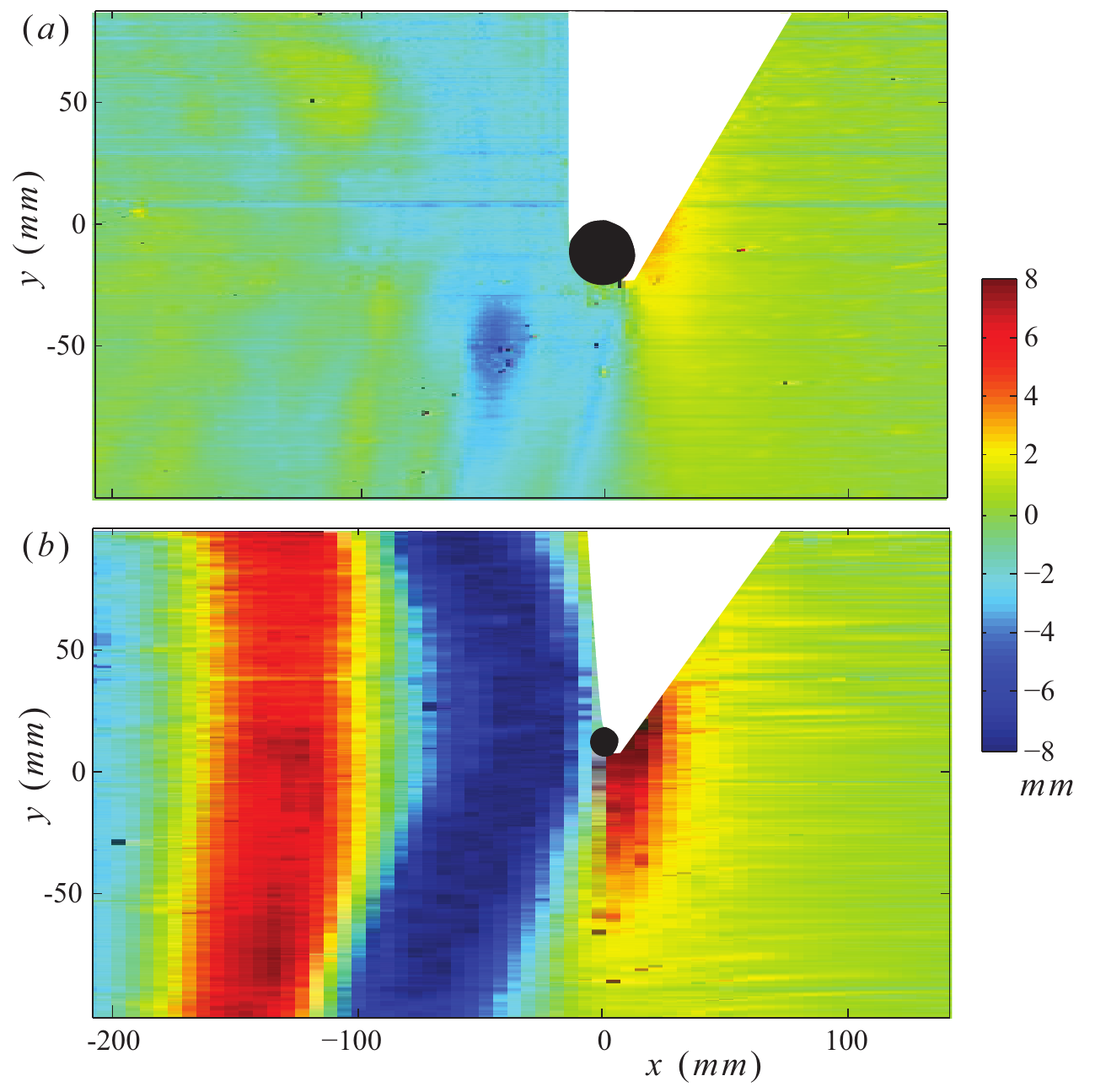}
\caption{Free-surface height reconstruction for $Re=8500$. (a) $D=3$ cm and $Fr=0.5$. (b) $D=1.4$ cm and $Fr=1.6$. The cylinders are indicated by black circles and the white bands correspond to shadows areas of the cylinders that cannot be analyzed.}
\label{fig::fig2}
\end{figure}

Figure \ref{fig::fig2} presents examples of free-surface reconstructions using the optical method for $Re=8500$. Two $Fr$ are shown: (a) $Fr=0.5$ and (b) $Fr=1.6$. When the $Fr$ increases, the free-surface experiences elevations of larger amplitude. The diagrams allow to distinguish the fountain height upstream of the cylinders. It is also possible to see the angle of the V-shaped wakes. For $Fr=1.6$ the V-shaped wake induces a planar wave downstream of the cylinder. Note that this method and other optical methods (e.g. Ref. \cite{phdgomit}) are restricted by the dispersion of the laser due to the cylinder turbulent wake. Hence, the present technique is limited for small $Fr$ and $Re$. 

\section{Results and discussion}

The results consist on drag measurements, surface elevation measurements and image analysis. These have been obtained for different cylinder diameters translated at various speeds with constant immersion $h$ and constant ratio $h/D$. The different modes of air-entrainment are depicted and used to detect the critical velocity for air-entrainment in the cavity. The scaling of the velocity can be explained using a simple model described below.

\subsection{Drag measurements}

\begin{figure}
\center
\includegraphics[width=0.8\columnwidth]{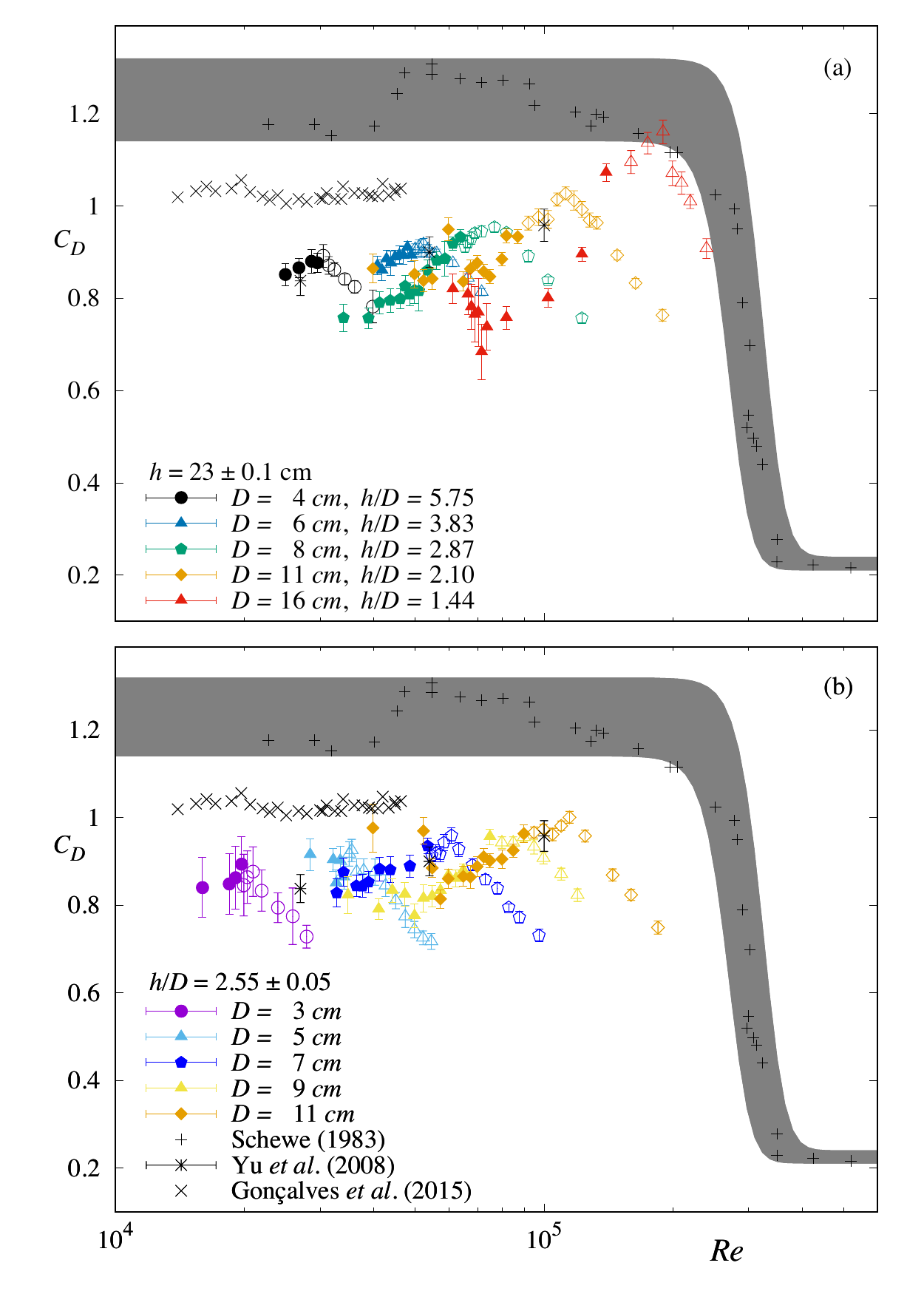}
\caption{Drag coefficients, $C_D$, as a function of Reynolds number for different cylinder diameters $D$. (a) Constant immersion depth $h=23\pm 0.1$ cm and (b) constant ratio $h/D=2.55 \pm 0.05$. Filled symbols represent situation without air-entrainment and empty symbols represent flow with air-entrainment behind the cylinder. Additional results from the literature are added and the results without free-surface are highlighted using a gray band.}
\label{fig:fig3}
\end{figure}

In this section, force measurements are reported in a range of $Re$ and $Fr$, such that the transition to air-entrainment is observed. Figure \ref{fig:fig3} presents drag coefficients measurements for different cylinder diameters as a function of $Re$, at constant $h=23\pm0.1$ cm and at constant ratio $h/D=2.55\pm0.05$. Filled (resp. empty) symbols correspond to cases without (resp. with) air-entrainment in the cavity $\mathcal{L}$ behind the cylinder. The error bars correspond to the standard deviation of the $C_D$. The results obtained are compared with results of Schewe \cite{schewe1983} for single phase flow. Additional results of Refs. \cite{yu2008,goncalves2015} on surface piercing cylinders are represented for the same range of $Re$ and $h/D$. In accordance with previous studies, the drag coefficient is found to be about 30\% smaller for surface piercing cylinders than for single phase flows. The numerical results of Yu {\it et al.} \cite{yu2008} at three different $Re$ and a constant ratio $h/D=4$ are in good agreement with the present measurements. The results of Gon\c{c}alves {\it et al.} \cite{goncalves2015} for a ratio of $h/D=2$ show a nearly constant drag coefficient at higher values than the present study. In their experiments the cylinder is attached to the walls of the flume and the aspect ratio $h/D$ is varied by varying the water depth. Moreover, their maximum $Fr$ based on the cylinder diameter is $Fr=0.36$, lower than the present values and below the threshold of air-entrainment. This may explain their highest $C_D$ values.

In the present study, the drag coefficient first increases with the $Re$ then decreases after air-entrainment occurred in the cavity downstream the cylinder. Air-entrainment is defined when the first bubble is observed or monitored from the camera images (i.e.  bubble larger than 300 $\mu$m). In figure \ref{fig:fig3}, for small $h/D$ or large diameters $D$, $C_D$ values first increase with $Re$ after the appearance of first air bubbles. Then, $C_D$ values decrease from 10\% for the 4 cm diameter cylinder to 26\% at maximum for the 11 cm diameter cylinder. The existence of drag reduction effect could be explained by the modification of pressure distribution along the cylinder. 

\begin{figure}
\center
\includegraphics[width=0.8\columnwidth]{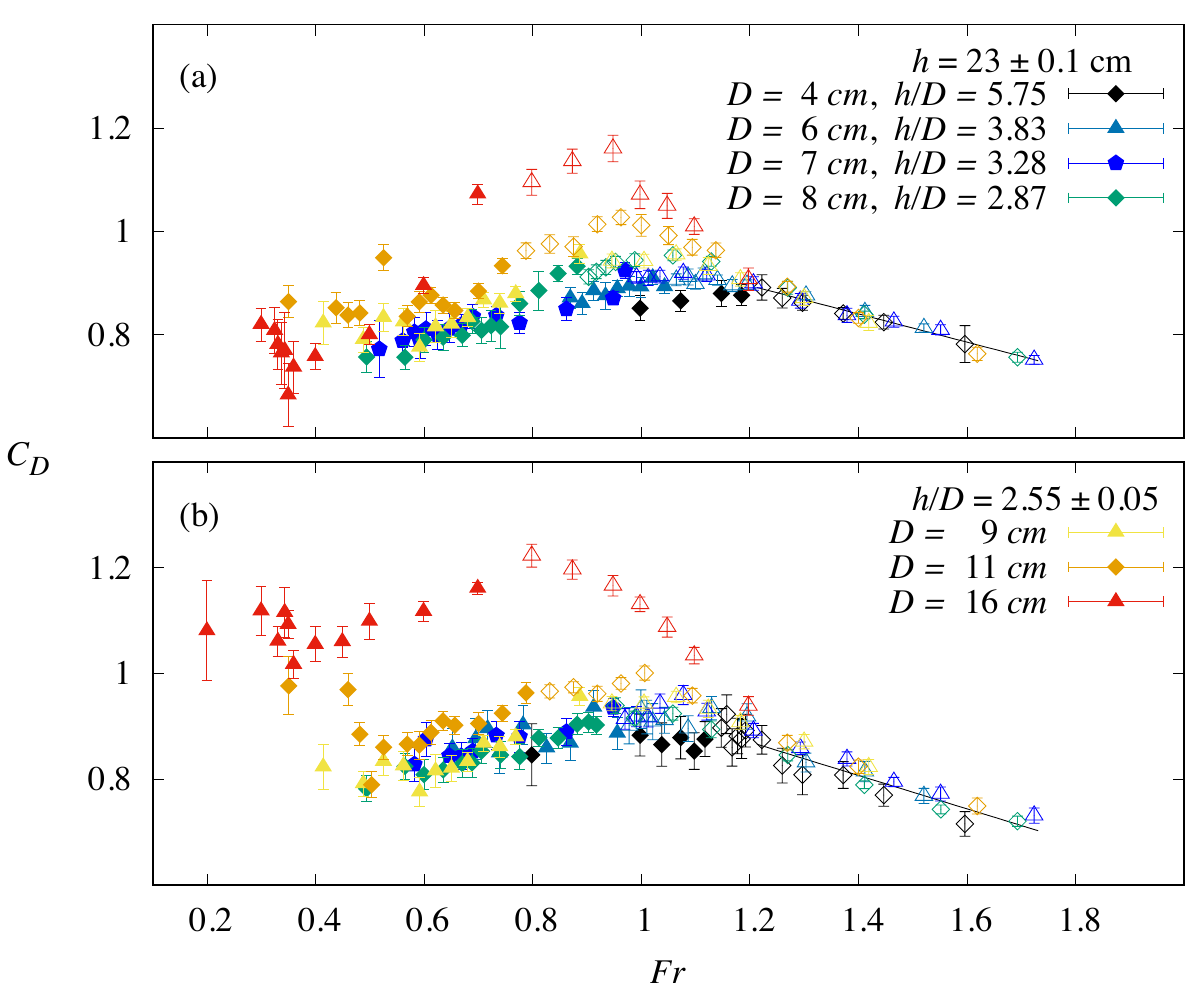}
\caption{Drag coefficients, $C_D$, as a function  of Froude number, $Fr$, for different cylinder diameters $D$. (a) corresponds to a constant immersion depth $h=23\pm 0.1$ cm  and (b) to a constant ratio $h/D=2.55\pm 0.05$. Filled symbols represent situation without air-entrainment and empty symbols represent flow with air-entrainment in the cavity. For $Fr>1.2$, $C_D$ is linear and the black line is a linear fit.}
\label{fig:fig4}
\end{figure}

In order to study the influence of gravity effect, the drag coefficient $C_D$ is also plotted as a function of $Fr$ for a constant immersion depth, $h=23\pm0.1$ cm, in Fig. \ref{fig:fig4}(a) and for constant ratio, $h/D= 2.55 \pm 0.05$, in Fig. \ref{fig:fig4}(b). As in Fig. \ref{fig:fig3}, filled (resp. empty) symbols correspond to measurements without (resp. with) air-entrainment in the cavity. In both cases, the evolution of the drag coefficient clearly shows a transition occurring at $Fr \,\approx\,1.2$. For $Fr<1.2$, $C_D$ is roughly larger for large $D$ or $D/h$. For $Fr>1.2$, after air-entrainment, $C_D$ values are monotonous and seem to vary linearly with $Fr$. A similar linear behavior has also been observed in granular flows around an immersed cylinder \cite{chehata2003}.

\subsection{Air-entrainment}

Two mechanisms of air-entrainment are revealed: (i) in the wake thanks to vortices and (ii) in a cavity behind the cylinder. For $1.4 <D< 4$ cm ($5\,200 < Re/Fr< 25\,000$), air-entrainment occurs in the cavity along  the cylinder exclusively. This cavity \cite{yu2008,koo2014} forms viscous cusps where air-entrainment is systematically injected. Once air bubbles form below the cavity, they are carried downstream into the wake. The bubbles created by air-entrainment in the cavity have an equivalent diameter ranging from 6 mm for $D=3$ cm to 8.5 mm for $D=16$ cm. 

\begin{figure}
\center
\includegraphics[width=1.0\columnwidth]{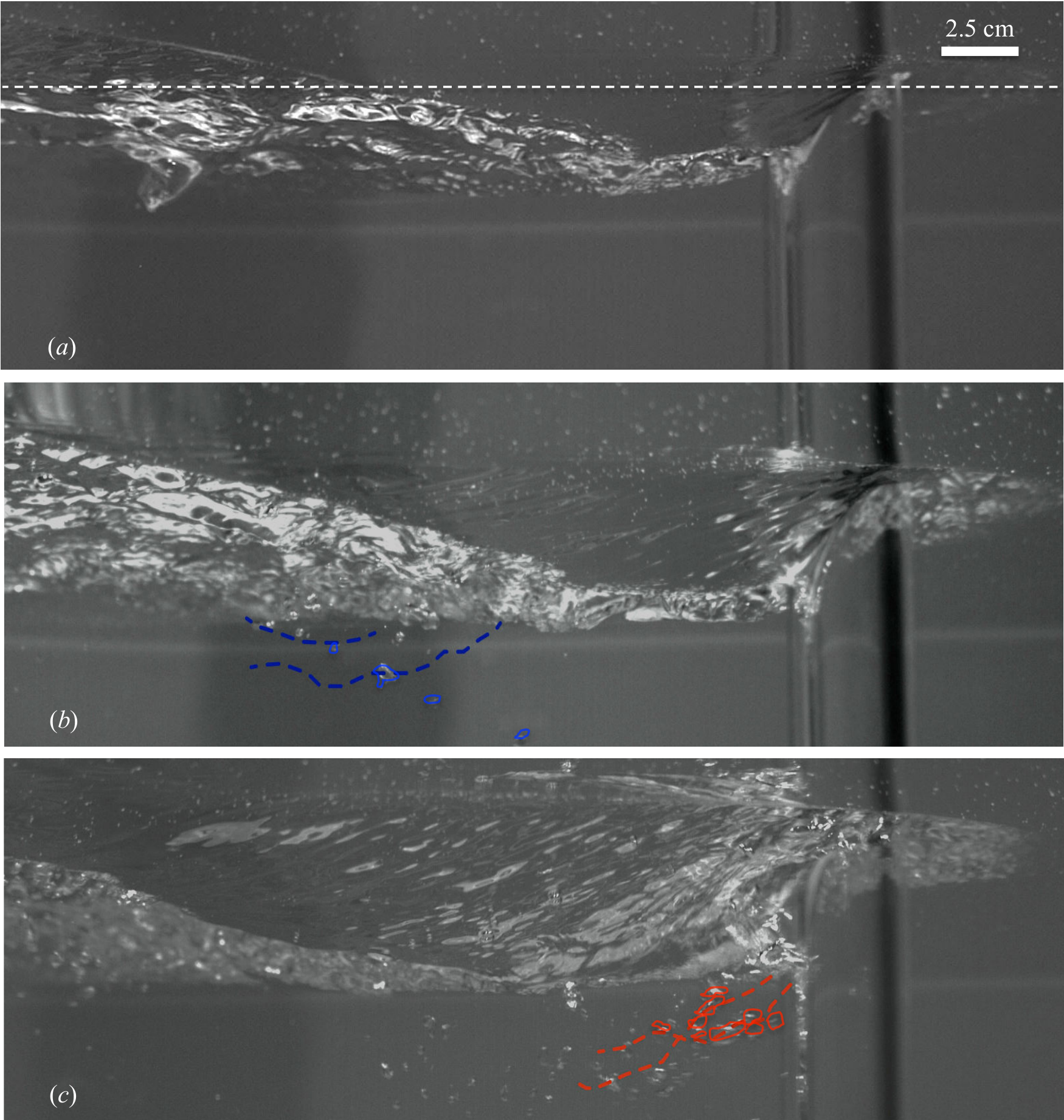}
\caption{View from the side below-surface of the cylinder of 5 cm diameter. (a) No air-entrainment at $Re = 34\, 700$ or $Fr=0.97$, see supplementary movie 1, (b) air-entrainment in the wake at $Re = 39\, 700$ or $Fr=1.11$, bubbles encircled in blue, see supplementary movie 2, (c) air-entrainment in the cavity behind the cylinder at $Re=45\,680$ or $Fr=1.3$, bubbles encircled in red, see supplementary movie 3. The dashed lines represent bubble trajectories.}
\label{fig:fig5}
\end{figure}

\begin{figure}
\center
\includegraphics[width=1.0\columnwidth]{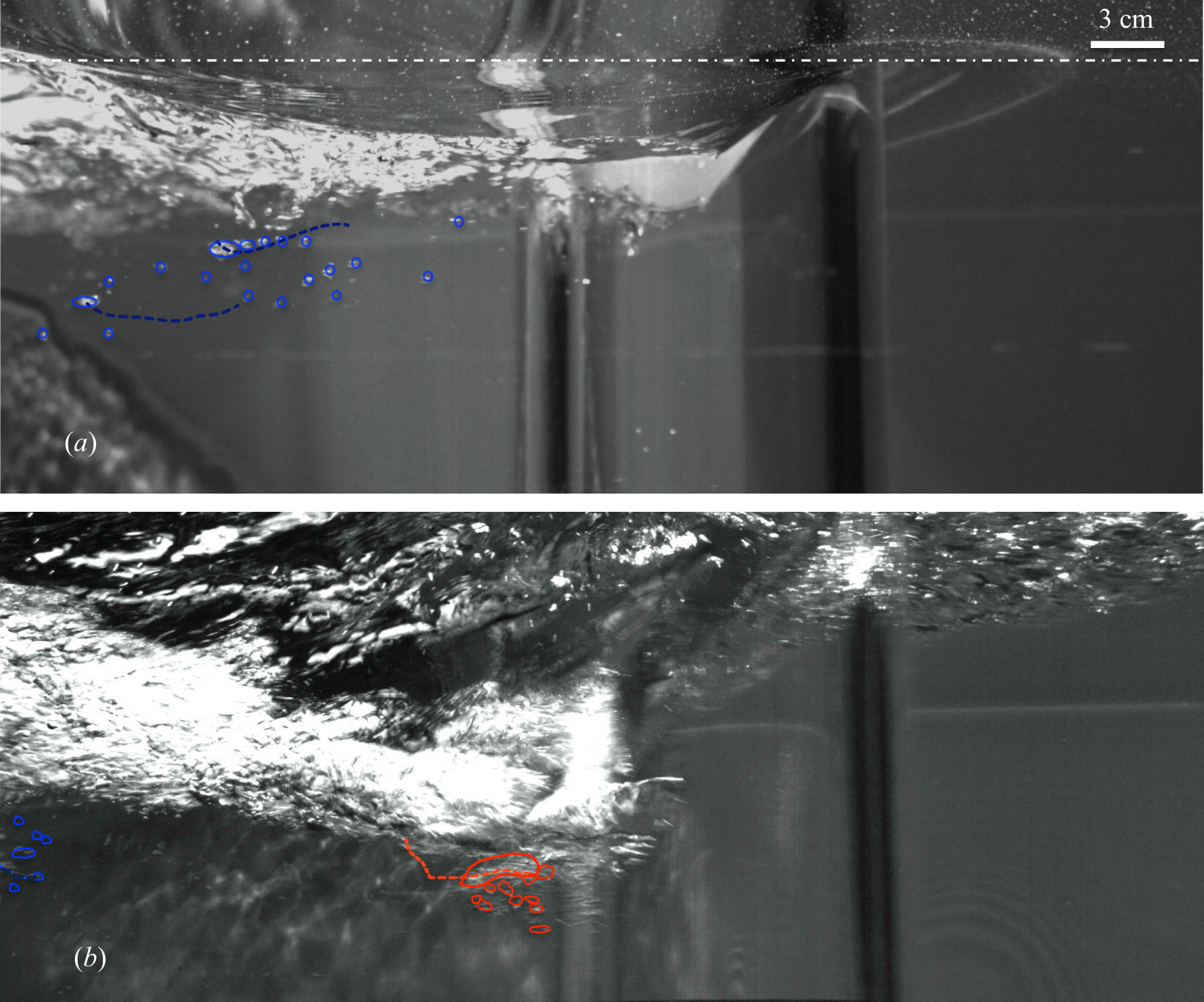}
\caption{View from the side below-surface of the cylinder of 16 cm diameter. (a) Air-entrainment in the wake at $Re = 122\, 500$ or $Fr=0.6$, bubbles encircled in blue, see supplementary movie 4. (b) Air-entrainment in the cavity behind the cylinder at $Re=199\,300$ or $Fr=1$, bubbles  encircled in red, see supplementary movie 5. The dashed lines represent observed bubble trajectories.}
\label{fig:fig6}
\end{figure}

For $4<D<16$ cm, air-entrainment first appears in the wake when $Re$ is increased. Then, as the velocity increases, air-entrainment also appears in the cavity. For a cylinder of 5 cm diameter, Fig. \ref{fig:fig5}(a), at $Re=34\, 670$, shows free-surface deformation without air-entrainment. In Fig. \ref{fig:fig5}(b) at $Re=39\, 700$, air-entrainment occurs in the cylinder wake. Finally, in Fig. \ref{fig:fig5}(c), at $Re=45\, 700$, the cavity induces air-entrainment more intensely. For $D=16$ cm and higher $Re$, the mechanisms of air-entrainment presented in Fig. \ref{fig:fig6} are similar. Air-entrainment appears in the wake, see Fig. \ref{fig:fig6}(a), then in the cavity, see Fig. \ref{fig:fig6}(b). The corresponding movies are available as supplementary materials.

\begin{figure}
\center
\includegraphics[width=0.8\columnwidth]{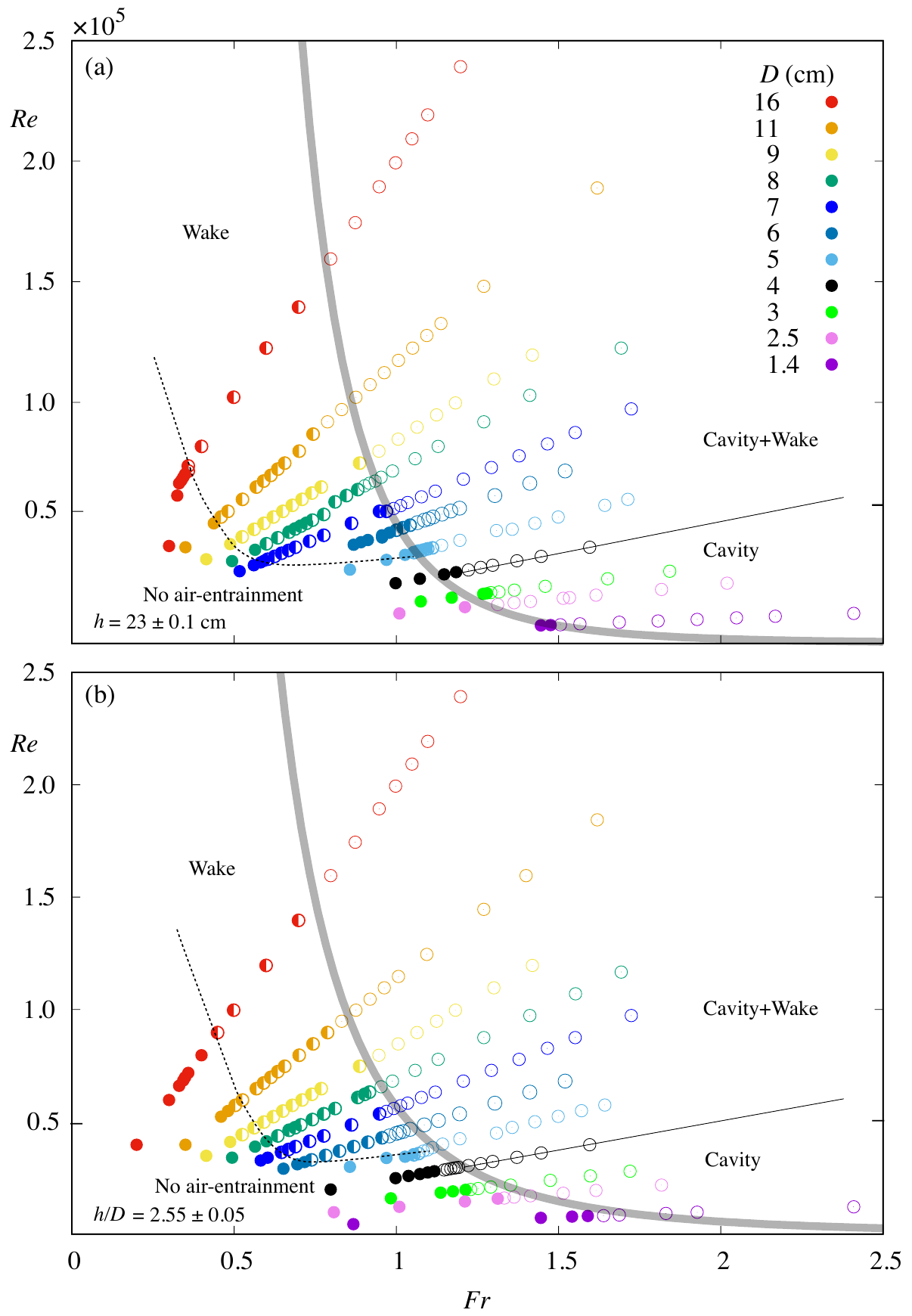}
\caption{Flow diagrams $Re$ versus $Fr$ depicting the different air entrainment regimes: no-air entrainment (filled symbols), air entrainment in the cavity (partially filled symbols) and in both the cavity and the wake (empty symbols) for different cylinder's diameters or $Re/Fr$. (a) For constant  immersion $h=23\pm0.1$ cm and (b) constant $h/D=2.55\pm0.05$. The lines are guides to the eyes delimiting the different air-entrainment regimes.}  
\label{fig:fig7}
\end{figure}

A summary of all our observations is presented in Fig. \ref{fig:fig7}, for both constant $h$ and $h/D$. In both cases, several regimes are observed where the different modes of air-entrainment are: (i) in the wake, (ii) in the cavity and (iii) in the wake and the cavity. Boundaries can be drawn between the different regimes and it is the purpose of the model below to explain the nature of some of these boundaries. The main difference between the two diagrams is the area of the wake region, which is wider for large diameters and constant $h$. For small $h/D$, the onset of air-entrainment occurs at smaller $Fr$ presumably because of the size of the recirculation below the free-surface \cite{benitz2016,goncalves2015}. Additionally, the fact that the two diagrams are almost indistinguishable suggests that the immersed height, $h$, has little effect on the air-entrainment when $h>23\pm0.1$ cm and $h/D>2.55\pm0.05$. Hence, air entrainment is mainly due to the free-surface dynamics and deformation.

\begin{figure}
\center
\includegraphics[width=1.0\columnwidth]{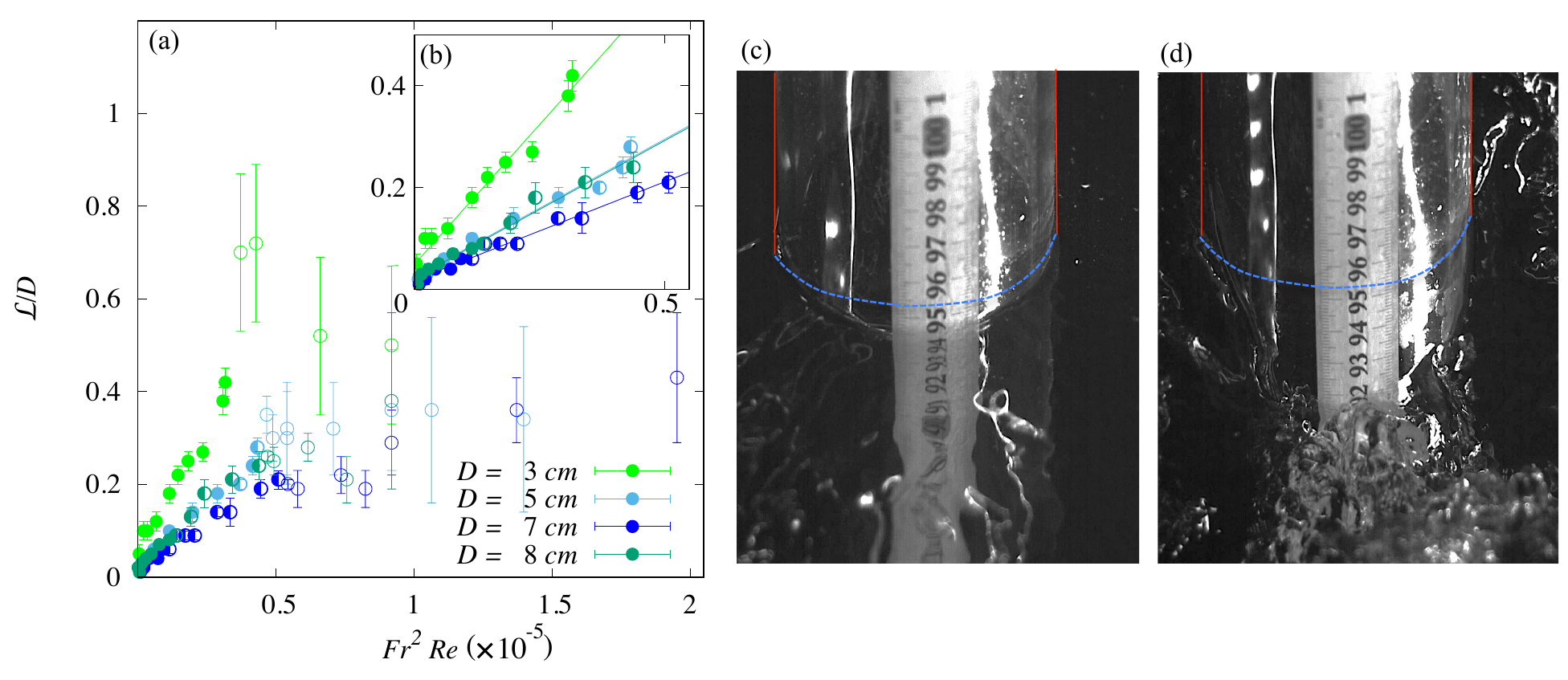}
\caption{(a) Non-dimensional cavity depth $\mathcal{L}/D$ as a function of $Fr^2 Re$, for different diameters. The inset (b) is a zoom restricted to the regime of no or weak air-entrainment, where the continuous lines are linear fits. Filled symbols represent situation without air-entrainment and hollow symbols represent flow with air-entrainment in the cavity.  (c) and (d) are snapshots of the cavity, view from the back, typical of no-air entrainment and air-entrainment, respectively. The continuous vertical red lines represent the emerged cylinder walls and the dashed bleue line represents the free-surface at rest.}  
\label{fig:fig8}
\end{figure}

Indeed, the main feature associated to air-entrainment is the deformation of the cavity behind the cylinder. The optical method described earlier and the direct observations of graduations along the cylinder allowed to measure the cavity depth $\mathcal{L}$ for various conditions. Specifically, the observations of graduated cylinders are reported in Fig. \ref{fig:fig8}: $\mathcal{L}/D$ as a function of $Fr^2 Re$. $\mathcal{L}$ represents the mean value from 10 snapshots (see Fig. \ref{fig:fig8}(c) and (d)). The error bars are the maximum observed deviation from the mean value. Clearly, the behavior is nonlinear, typically after air-entrainment takes place. However, in the range of $Fr^2 Re$ before air-entrainment, $\mathcal{L}/D$ varies linearly with $Fr^2 Re$. The proportionality coefficients seem to decrease with the diameter. These measurements are insufficient to provide a definite scaling of the cavity height. In fact, the cavity is unsteady and the liquid motion is more disordered as the velocity increases.

Air-entrainment phenomenon in the cavity could be explained using the model developed by Benusiglio \cite{benusiglio2013} where the acceleration of the cavity $\gamma$ is balanced by the gravitational acceleration $g$. $\gamma$ is controlled by the vortex shedding frequency $f$, calculated from the Strouhal number. In the range of $Re$ of the present study, $St$ is assumed constant ($\sim 0.2$) \cite{schewe1983} although some studies have shown its variability upon the cylinder aspect ratio \cite{farivar1981,fox1993,benitz2016}. The acceleration $\gamma$ of the cavity of depth $\mathcal{L}$ is defined as follows:
\begin{equation}
\gamma = \mathcal{L} f^2.
\label{eq2}
\end{equation}
In equation (\ref{eq2}), $\mathcal{L}$ is determined using Bernoulli's equation between two positions upstream and downstream of the cylinder. At upstream position the mean flow velocity is $U$ whereas at the downstream position the flow velocity is taken as the maximum velocity $\mathcal{U}$ of a Rankine vortex ${\Gamma}/{2\pi a}$ \cite{li2008} where $\Gamma$ is the circulation around half of the cylinder ${DU}/{2}$, and $a$ represents the core radius of the vortex. The parameter $a$ scales with ${D}/{\sqrt{Re}}$ the boundary layer thickness \cite{schlichting1955}. With these assumptions, Benusiglio \cite{benusiglio2013} proposed a scaling for $\mathcal{L}$:
\begin{equation} 
\mathcal{L} \propto \frac{DU^{3}}{g \nu},
\label{eq3}
\end{equation}
which has not been validated experimentally or numerically. However, our current measurements, see Fig. \ref{fig:fig8}, using the direct cavity depth observations, seem to be consistent with $\mathcal{L}/D \propto Fr^2 Re$, which is equivalent to equation (\ref{eq3}). Finally, air-entrainment occurs when $\gamma \simeq g$, therefore equation (\ref{eq2}) leads to the critical velocity $U_c$ for air-entrainment in the cavity $\mathcal{L}$:
\begin{equation} 
U_c \propto (\nu g^{2})^{\frac{1}{5}} D^{\frac{1}{5}}
\label{eq4}
\end{equation}

\begin{figure}
\center
\includegraphics[width=1.0\columnwidth]{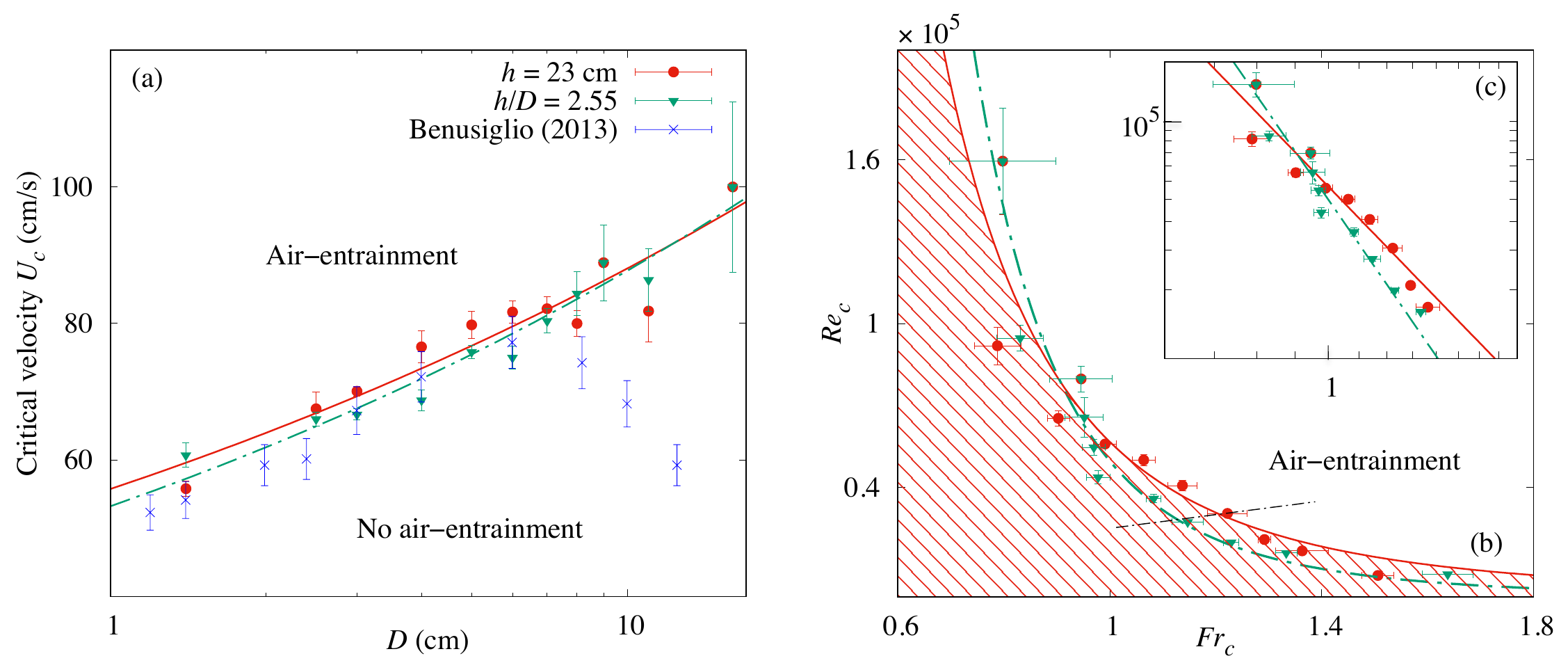}
\caption{(a) Critical velocity $U_c$ for air-entrainment in the cavity $\mathcal{L}$ as a function of the cylinder diameter $D$ for constant  immersion $h=23\pm0.1$ cm and constant $h/D \simeq 2.55\pm0.05$. Results of Benusiglio \cite{benusiglio2013} are plotted with blue crosses. Green dotted line represents power law ${D^{ \,0.22 \pm 0.02}}$ for the constant $h/D$ case and red line power law $D^{\, 0.20 \pm 0.02}$ for $h=23\pm0.1$ cm. The shaded region corresponds to no air-entrainment. (b) Critical Reynolds number, $Re_c$, as a function of the critical Froude number, $Fr_c$. The black dashed-doted line represents $Re/Fr \simeq 25\,000$. The inset (c) is the same data in log-log scale.}  
\label{fig:fig9}
\end{figure}

Figure \ref{fig:fig9}(a) shows the critical velocity $U_c$ for  air-entrainment in the cavity $\mathcal{L}$ as a function of $D$ at constant immersion $h=23\pm 0.1$ cm and constant $h/D=2.55 \pm 0.05$. The error bars of $U_c$ correspond to the velocity step between the maximum velocity when air-entrainment is not obseved and the velocity where the first bubble appears. The evolution of $U_c$ is fitted with power laws of $D^{\,0.2 \pm 0.02}$ for $h/D \simeq 2.55$ and $D^{\,0.22 \pm 0.02}$ for $h \simeq 23$ cm. Experimental results show good agreements with the theoretical model introduced with equation (\ref{eq4}). The simple model based on energy conservation with Bernoulli's equation predicts the appearance of air-entrainment in the cavity. Additional data from Benusiglio \cite{benusiglio2013} are also plotted and fit well the present data for  $D \le 6$ cm. For larger diameters, the critical velocity given by Benusiglio \cite{benusiglio2013} refers to air-entrainment in the cylinder wake like in Fig. \ref{fig:fig5}($b$), while our present data refers only to air-entrainment in the cavity. Figure \ref{fig:fig9}($b$) presents the evolution of $Re_c$ as a function of $Fr_c$, based on the critical velocity $U_c$. The dimensionless form of the equation (\ref{eq4}) is $Re_c \propto Fr_c^{-4}$. The evolution of $Re_c$ is fitted with a power law of $Fr_c^{\,-3.3 \pm 0.62}$ for $h=23$ $\pm$ 0.1 cm and $Fr_c^{\,-4.7 \pm 0.51}$ for $h/D=$ 2.55 $\pm$ 0.05, in good agreement with the prediction of the model. The red shaded area refers to the range of dimensionless parameters where no air-entrainment in the cavity occurs. The black dashed-doted line represents the threshold $Re/Fr \simeq 25\,000$ above which air-entrainment occurs in the wake and in the cavity. Hence, air-entrainment requires large inertial effects or strong gravity effects. It is interesting to note that the stability diagram in Fig. \ref{fig:fig9}($b$) is consistent with the results of Kumagai {\it et al.} \cite{kumagai2011} for horizontal cylinder translating beneath an air-water interface.

\section{Conclusions}

Experimental results on the dynamics of the flow around a vertical translated cylinder have been presented. These deal with the turbulent regime before the so-called fully ventilated regime when air reaches the bottom of the cylinder. There is a critical velocity for the onset of air-entrainement. Then, three modes of air-entrainment are observed depending on the diameter and the translating velocity. For $D \leq4$ cm, air-entrainment only occurs in the cavity downstream of the cylinder. For $D\geq 5$ cm, air-entrainment first appears in the wake and then also in the cavity when $Re$ or $Fr$ is increased. This study focuses on air-entrainment in the cavity, where forces, flow visualization and critical velocities for air-entrainment  are reported at constant $h$ or $h/D$. Drag force measurements show an increase of the drag coefficient with $Re$ followed by a decrease after air-entrainment occurred in the cavity. A similar evolution is observed for $C_D$ versus $Fr$. Moreover, for $Fr > 1.2$, the drag for all diameters overlap and vary linearly with $Fr$. The threshold velocity for air-entrainment in the cavity scales as: $U_c \propto D^{\, 0.2}$, i.e $Re_c\propto Fr_c^{-4}$, which is in good agreement with the scaling of the model of Benusiglio \cite{benusiglio2013}. In the future, the shape of the cavity will be studied in more details in order to quantify the contribution of wetting properties \cite{aristoff2009,truscott2014} and the capillary-gravity waves.

\begin{acknowledgments}
The authors acknowledge the financial support of the Normandy council and European Union ERDF funding's through the NEPTUNE project. We thank Antoine Bonnesoeur and Claude Houssin for his engineering support and Paul Fran\c{c}ois for his help with the experiments. Our work has also benefited from exchanges with Yuji Tasaka (U. Hokkaido) and Jimmy Philip (U. Melbourne).
\end{acknowledgments}

\bibliographystyle{unsrt}
\bibliography{biblioCylinder}

\end{document}